\documentclass[aps,twocolumn]{revtex4}
\usepackage{graphicx}
\usepackage{amsfonts}
\usepackage{amssymb}
\usepackage{amsbsy}
\usepackage{amsmath}
\usepackage{mathrsfs}
\usepackage{latexsym}
\usepackage{natbib}
\usepackage{bm}
\usepackage{color}
\usepackage{braket}
\usepackage{slashed}
\usepackage[hidelinks]{hyperref}
\usepackage[all]{hypcap}
\usepackage[normalem]{ulem}

\def\k{\mathrm{k}}
\def\gf{\mathcal{G}}
\def\am{\mathcal{A}}

\begin{document}

\author{Golam Mortuza Hossain}
\email{ghossain@iiserkol.ac.in}

\author{Pushpit Kumar}
\email{pk21ms209@iiserkol.ac.in}

\affiliation{ Department of Physical Sciences,
Indian Institute of Science Education and Research Kolkata,
Mohanpur - 741 246, WB, India }
	

\title{Neutrino Oscillations without Mass: A Re-analysis of KamLAND Data
following the Dirac Equation in Curved Spacetime}
	
\begin{abstract}
The Dirac equation in stationary curved spacetime implies that describing
two-flavor neutrino oscillations requires treating the \emph{invariant} mass
and \emph{conserved} energy of each propagating state distinctly, as gravity
affects energy differently than mass. By re-analysing the publicly
released KamLAND dataset, we show that modeling flavor states of massless
neutrinos as two-level quantum states with energy-dependent level splitting,
analogous to modified Jaynes-Cummings models, results in a higher-likelihood fit
to the observed data than the standard massive-neutrino paradigm.
\end{abstract}

\maketitle

\subsection{Introduction}

In the Standard Model of particle physics, neutrinos are treated as massless
elementary particles that occur in three flavors, namely the electron neutrino
$\nu_e$, the muon neutrino $\nu_{\mu}$, and the tau neutrino $\nu_{\tau}$
\cite{navas2024review}. However, various experiments studying solar,
atmospheric, and reactor neutrinos \cite{davis,fukuda1998evidence,
ahmad2002direct, kamlandFirst2003} have shown that neutrinos produced in a given
flavor state can be detected later in a different flavor state.

This conversion phenomenon is usually understood in terms of the flavor
\emph{mixing} where flavor states are treated as a coherent superposition of the
mass eigenstates. In particular, flavor conversion of reactor antineutrinos,
such as those observed by the Kamioka Liquid-scintillator Antineutrino Detector
(KamLAND) experiment \cite{kamlandFirst2003, kamlandSecond2005, Kamland2008}, is
explained using neutrino oscillations \cite{Pontecorvo, maki1962remarks}, where
an antineutrino produced in one flavor state propagates over long baselines as a
superposition of mass eigenstates. Solar neutrino observations, on the other
hand, are understood primarily in terms of adiabatic flavor conversion through
the Mikheyev-Smirnov-Wolfenstein (MSW) effect \cite{Wolfenstein,MSW}. In these
frameworks, the neutrinos are treated as massive particles, which in turn
necessitates physics beyond the Standard Model.

The detection of flavor neutrinos from supernova SN1987A
\cite{KamiokandeSN1987A, BaksanSN1987A, IMBSN1987A} resulting to a neutron star
core \cite{SN1987Ascience.adj5796}, nevertheless, raises question on the notion
of flavor states in massive-neutrino paradigm, due to the loss of coherence
over astrophysical distances \cite{Kersten_2016, TUREANU2023137996,
TUREANU2025117052}.
Recently, neutrino oscillations in the curved spacetime of neutron stars have
been studied \cite{bandyopadhyay2025probing} by employing the Dirac equation in
curved spacetime as a probe of possible equations of state
\cite{hossain2021equation, hossain2021higher,hossain2022equation} which
incorporate the effects of extreme gravity. This generally covariant approach
implies that two-flavor neutrino oscillations are controlled by four distinct
parameters: the invariant mass and conserved energy of each propagating state
where gravitational time-dilation impacts energy differently than mass. This
suggests that one should re-examine the role of the energy difference
independently from the usual mass-squared difference, unlike standard analysis
in flat spacetime. The KamLAND dataset \cite{kamlandR2Data}, which provides
baseline-dependent oscillation data, is particularly well-suited for performing
such a re-analysis.

\subsection{Conserved energy of a particle}
\label{sec:Conserved_enegy_of_a_particle}

In curved spacetime, the energy of a particle is not a conserved quantity in
general. However, if the spacetime admits a timelike vector field $t^{\mu}$
that satisfies the Killing equation $\nabla_{(\mu} t_{\nu)} = 0$, then
for a particle following a geodesic with an affine parameter $\lambda$ and
four-momentum $p^{\mu}$, one can define a conserved energy $\varepsilon
\equiv - t^{\mu} p_{\mu}$ since it satisfies
\begin{equation} \label{EnergyConservationEq}
\frac{d\varepsilon}{d\lambda} = p^{\nu} \nabla_{\nu} \varepsilon
= - p^{\nu} p^{\mu} \nabla_{\nu} t_{\mu}
- t^{\mu} p^{\nu} \nabla_{\nu} p_{\mu}  = 0 ~,
\end{equation}
due to the Killing equation and the geodesic equation $p^{\nu} \nabla_{\nu}
p_{\mu} = 0$. This result holds for both massive and massless particles.

\subsection{Spherically Symmetric Spacetime}
\label{sec:Spacetime_metric}

The spacetime around a gravitating body such as the Earth or a star may be
described by a spherically symmetric metric. In \emph{natural} units
($c=\hbar=1$), the corresponding invariant distance element is given by
\begin{equation} \label{SphericallySymmetricMetric}
ds^2 = -e^{2\Phi} dt^2 + e^{2\nu} dr^2
+ r^2 \left( d\theta^2 + \sin^2\theta\, d\varphi^2 \right) ,
\end{equation}
where $\Phi = \Phi(r)$ and $\nu = \nu(r)$ are the metric functions. In the
exterior region, they have exact solutions of the form $e^{2\Phi} = e^{-2\nu}
= 1 - 2GM/r$ where $G$ is Newton’s constant for gravitation and $M$ is the mass
of the spherical body.

\subsection{Fermions in curved spacetime}
\label{sec:Fermions_in_curved_spacetime}

Here we consider the neutrinos to be Dirac fermions whose dynamics in a curved
spacetime, using the Fock-Weyl formulation, is governed by the generally
covariant Dirac action \cite{fock1929geometrisierung, weyl1929electron} given by
\begin{equation}\label{DiracActionCurved}
S_\psi = - \int d^4x\, \sqrt{-g}\,
\bar{\psi}\left[i\gamma^a e^\mu_{\;a} \mathcal{D}_\mu + m\right]\psi ,
\end{equation}
where $\bar{\psi}=\psi^\dagger\gamma^0$ denotes the Dirac adjoint and $m$ is the
mass of the fermion. The tetrads $e^\mu_{\;a}$ relate the curved spacetime
metric with the local Minkowski frame through the relation $g_{\mu\nu}
e^\mu_{\;a} e^\nu_{\;b} = \eta_{ab}$ with $\eta_{ab}=\mathrm{diag}(-1,1,1,1)$.
The spin-covariant derivative $\mathcal{D}_\mu$ is defined as $\mathcal{D}_\mu
\equiv \partial_\mu + \Gamma_\mu$ with $\Gamma_\mu$ being the spin
connection given by
%
$\Gamma_\mu = -\frac{1}{8} \eta_{ac} {e_{\nu}}^c (\partial_\mu
{e^{\nu}}_b + \Gamma^\nu_{\mu\sigma} {e^{\sigma}}_b ) [\gamma^a,\gamma^b] ~,
$
%
where $\Gamma^\nu_{\mu \sigma}$ are the Christoffel connections and $\gamma^a$
are the Dirac matrices in Minkowski spacetime, satisfying the Clifford algebra
$\{\gamma^a, \gamma^b\} = -2 \eta^{ab} \mathbb{I}$.

In a spherically symmetric spacetime (\ref{SphericallySymmetricMetric}),
the Dirac action (\ref{DiracActionCurved}) implies that the \emph{radial}
modes with wave-vector $\k_r$ satisfy the equation $i\partial_t \psi_{\k} =
H^{0}_{\k} \psi_{\k} = \varepsilon_{\k} \psi_{\k}$
\cite{bandyopadhyay2025probing} where the eigenvalues of the mode Hamiltonian
$H^{0}_{\k}$ are
\begin{equation}
\label{ModeEnergyEigenvalue}
\varepsilon_{\k} = e^{\Phi} \sqrt{(e^{-\nu} \k_{r})^2 + m^2}  ~.
\end{equation}
To show that these eigenvalues are conserved, we note that the spacetime
(\ref{SphericallySymmetricMetric}) admits a timelike Killing vector
$t^{\mu} = (1,0,0,0)$. If we identify the wave-vector $\k_r$ with the
radial co-momentum $p_r = e^{2\nu} p^r$ where $p^r = m u^r = (mv)u^0$ with
$v=(dr/dt)$ then the eigenvalues (\ref{ModeEnergyEigenvalue}) have the form
of the conserved energy $\varepsilon = e^{\Phi} \sqrt{(e^{\nu} p^{r})^2 + m^2}$
which follows from (\ref{EnergyConservationEq}).
In the limit $\Phi\to 0$, $\nu\to 0$, $\varepsilon$ reduces to its Minkowskian
expression as expected. In the Newtonian non-relativistic limit \emph{i.e.}
$(GM/r) \ll 1$ and $v^2 \ll 1$ it becomes $\varepsilon \simeq m  + \frac{1}{2} m
v^2 - G M m/r$  in the exterior of the spherical body. Thus, the quantity
$\varepsilon$ indeed represents the total energy of a particle including its
rest energy $m$ to an asymptotic observer. Equation (\ref{ModeEnergyEigenvalue})
also implies that while the energy $\varepsilon$ is conserved, the spatial
momentum is \emph{not} a conserved quantity in a curved spacetime.

\subsection{Neutrino oscillations}

The results of the KamLAND experiment \cite{kamlandFirst2003, kamlandSecond2005}
are well described within a two-flavor neutrino oscillations framework where the
MSW effect is small. We denote the energy eigenstates as $\psi_1(t)$ and
$\psi_2(t)$. The flavor states are then expressed as $\bar{\nu}_{e}(t) =
\cos\theta \psi_1(t) + \sin\theta \psi_2(t)$ and $\bar{\nu}_{\beta}(t) = -
\sin\theta \psi_1(t) + \cos\theta \psi_2(t)$ where $\beta$ may refer to muon or
tau neutrinos and $\theta$ is known as the vacuum mixing angle. Consequently,
the time-evolution equation for these flavor states can be written as
\begin{equation}\label{FlavourTimeEvolution}
i \partial_t \begin{pmatrix} \bar{\nu}_e \\ \bar{\nu}_{\beta} \end{pmatrix}
= \left[ \varepsilon ~\mathbb{I} - \frac{\Delta\varepsilon}{2}
\begin{pmatrix}
  -\cos2\theta  & \sin2\theta \\
 \phantom{-}\sin2\theta  & \cos2\theta
\end{pmatrix} \right]
\begin{pmatrix} \bar{\nu}_e \\ \bar{\nu}_{\beta} \end{pmatrix}   ~,
\end{equation}
where $\varepsilon = \tfrac{1}{2}(\varepsilon_1 + \varepsilon_2)$ and
$\Delta\varepsilon = (\varepsilon_1 - \varepsilon_2)$. Equation
(\ref{FlavourTimeEvolution}) shows that a non-zero mixing angle $\theta$ and
energy gap $\Delta\varepsilon$ leads to flavor oscillations over time. In this
context, the oscillations occur predominantly in the $1$–$2$ sector and so the
mixing angle $\theta$ is identified with $\theta_{12}$. The survival probability
of an electron antineutrino after propagating over a baseline can be expressed
as
\begin{equation} \label{SurvivalProbability}
P_{ee} = 1-\sin^2 2\theta_{12}\,
\sin^2\!\left(\frac{\Delta\phi_{{\rm osc}}}{2}\right) ~,
\end{equation}
where phase difference $\Delta \phi_{\rm osc} = \phi_1 - \phi_2$ with $\phi_j
= \int dt ~\varepsilon_j$.
In the KamLAND experiment, the detector and the reactors are located at fixed
spatial positions. So for a given baseline $L$ we may express the phase
difference as $\Delta \phi_{\rm osc} = \int_{0}^{L} dl ~\gf$ where the
\emph{gap function} $\gf$ is
\begin{equation} \label{GapgPhiDef}
\gf \equiv  \frac{(e^{-\Phi} \varepsilon_1)^2}{\sqrt{(e^{-\Phi}
\varepsilon_1)^2 - m_1^2}}
- \frac{(e^{-\Phi} \varepsilon_2)^2}{\sqrt{(e^{-\Phi}
\varepsilon_2)^2 - m_2^2}}  ~,
\end{equation}
and $dl$ is the infinitesimal proper spatial length along the neutrino
trajectory, defined as $dl^2 = \tilde{q}_{ab}dx^a dx^b$ with spatial metric
$\tilde{q}_{ab}$ and $a,b \in (r,\theta,\phi)$. In the equation
(\ref{GapgPhiDef}), we have used $m_j (dt/d\tau)_j = e^{-2\Phi} \varepsilon_j$
and the spatial momentum magnitude as $m_j (dl/d\tau)_j = \sqrt{(e^{-\Phi}
\varepsilon_j)^2 - m_j^2}$. Equation (\ref{GapgPhiDef}) holds true for a
general trajectory, unlike the equation (\ref{ModeEnergyEigenvalue}) which
assumes a radial motion.

In curved spacetime, the function $\gf$, consequently $\Delta \phi_{\rm osc}$,
depends on four distinct parameters $m_1$, $m_2$, $\varepsilon_1$ and
$\varepsilon_2$ where the gravitational \emph{time-dilation} factor $e^{\Phi}$
affects energy and mass terms differently. In the flat spacetime limit, for
massive neutrinos, under either the equal spatial-momentum approximation
(though with a factor of $2$ discrepancy; see \cite{bandyopadhyay2025probing})
or the equal energy approximation \cite{OKUN2003443,E1=E2}, the gap function
reduces to the standard expression $\gf = \Delta m^2_{12} / (2\varepsilon) +
\mathcal{O}(\varepsilon^{-2})$ with $\Delta m^2_{12} = m_1^2 - m_2^2$ and
$\varepsilon = \tfrac{1}{2} (\varepsilon_1 + \varepsilon_2)$. However, here we
aim to perform an analysis where neutrinos are massless \emph{i.e.} $m_1\to0,
m_2\to0$. In such cases, neutrino oscillations are possible provided gap
function $\gf$ remains non-vanishing with  $\varepsilon_1 \ne \varepsilon_2$.

\subsubsection{Ansatz for gap function $\gf$}
\label{subsec:GapFunction}

The action (\ref{DiracActionCurved}) implies that neutrinos in curved spacetime
experience non-trivial gravitational coupling via the spin connection
$\Gamma_{\mu}$ (see, e.g., \cite{hossain2022equation, hossain2024origin}).
Additionally, neutrinos interact with background matter through electroweak
interactions. In view of these interactions, and by analogy with two-level
quantum systems described by modified Jaynes-Cummings models \cite{BUCK1981132,
PhysRevA.45.6816}, we parametrize the gap function (\ref{GapgPhiDef}) as
\begin{equation} \label{GapgPhiAnsatz}
\gf = \frac{\am \, \varepsilon_s^{q-1}}{\varepsilon^{q}} ~,
\end{equation}
where $\am$ and $q$ are two positive parameters. For later analysis, we set
the energy scale $\varepsilon_s = 1 ~\textrm{MeV}$. The choice of $\am = (\Delta
m^2_{12}/2)$ and $q=1$ corresponds to the standard massive-neutrino paradigm,
whereas for values $q\ne1$, the gap function $\gf$ may be viewed as level
splitting of two-level quantum systems representing flavor states of massless
neutrinos.


\subsection{Re-analysis of KamLAND data}

In the KamLAND experiment, an electron antineutrino $\bar{\nu}_e$, with energy
$\varepsilon$, is detected through the inverse beta decay reaction, $\bar{\nu}_e
+ p \rightarrow e^+ + n$, when it collides with a proton in the scintillator
medium. The produced positron $e^+$, with total energy $\varepsilon^{+}$,
subsequently annihilates with an electron, emitting two gamma-ray photons.
Neglecting the kinetic energy of the proton, the neutron, and the electron, the
total energy deposited due to the positron annihilation, referred to as the
\emph{true} energy, can be written as $\varepsilon_t = \varepsilon^{+} + m_e$,
whereas the energy of the antineutrino can be expressed as
\begin{equation} \label{eq:Relation_neutrino_true}
	\varepsilon = \varepsilon_t + (m_n - m_p - m_e) ~,
\end{equation}
where $m_p$, $m_n$, and $m_e$ denote the masses of the proton, the neutron, and
the electron respectively. This process generates a prompt burst of
scintillation light which is measured by the detector and referred to
as the \emph{prompt} energy $\varepsilon_p$. To include the uncertainty in the
measurement process, we consider the energy-resolution function of the
detector $R(\varepsilon_p,\varepsilon_t)$ to be a Gaussian with the mean
$\varepsilon_p$ and a relative resolution of $0.07/\sqrt{\varepsilon_p
[\textrm{MeV}]}$ \cite{kamlandR2Data}.

Further, the detector is located around $1$ km below the Earth's surface whereas
the reactors are located at the surface of the Earth. At Earth's surface,
the metric function $\Phi \approx -10^{-9}$ and its variation along the baseline
is even smaller. Henceforth, we set $e^{\Phi} \simeq 1$.

\subsubsection{Extended unbinned $\chi^2$ analysis}
\label{subsec:Chi2Analysis}

In our re-analysis, we use three datasets provided by the second KamLAND data
release \cite{kamlandR2Data}, namely (i) list of prompt energies $\varepsilon_p$
for the total number of observed antineutrino events $N_{\mathrm{obs}}=258$,
(ii) estimated background rate, denoted as $(dN_{\mathrm{bg}}/d\varepsilon_p)$,
for \mbox{$^8$He/$^9$Li} events, accidental coincidences, and
${}^{13}\mathrm{C}(\alpha,n){}^{16}\mathrm{O}$ events, and (iii) the integrated
antineutrino flux at the Kamioka site, denoted as $\phi_{l,i}$, as a function of
binned baselines, labelled by $l$, and different isotopes, labelled by $i$,
respectively due to the four fissile isotopes, namely ${}^{235}\mathrm{U}$,
${}^{238}\mathrm{U}$, ${}^{239}\mathrm{Pu}$, and ${}^{241}\mathrm{Pu}$. We
consider the prompt energy window to be $2.6$ to $8.0~\mathrm{MeV}$, following
the KamLAND analysis \cite{kamlandSecond2005}.

In order to extract the oscillation parameters, we compare the observed data
with the theoretically expected event rate, defined as
\begin{equation} \label{Eq:TheoreticalRate}
\frac{dN_{\mathrm{th}}}{d\varepsilon_p} =
\mathcal{N} \int_{-\infty}^{\infty} d\varepsilon_t\,
R(\varepsilon_p,\varepsilon_t)\, \sigma(\varepsilon)\,
\sum_{l,i} P^l_{ee}  \phi_{l,i} \lambda_i(\varepsilon) ~.
\end{equation}
The normalization constant $\mathcal{N}$ is fixed by equating the total number
of expected events to $365.2$, for the no-oscillation case \emph{i.e.} with
$P^l_{ee}=1$ \cite{kamlandSecond2005}. We numerically integrate the resolution
function $R(\varepsilon_p,\varepsilon_t)$ up to six standard deviations around
the mean. We consider the inverse beta-decay cross section to be
$\sigma(\varepsilon) \propto \varepsilon^{+} \sqrt{\varepsilon^{+2} - m_e^2}$
~\cite{IBD}.
Following \cite{kamlandR2Data}, the reactor antineutrino spectra
$\lambda_i(\varepsilon)$ are taken from \cite{U235} for ${}^{235}\mathrm{U}$,
\cite{U238} for ${}^{238}\mathrm{U}$,
\cite{Pu239_241} for ${}^{239}\mathrm{Pu}$ and ${}^{241}\mathrm{Pu}$.

We extract the best-fit oscillation parameters by using an extended
unbinned $\chi^2$ function, defined as
\begin{eqnarray}\label{eq:Chi_Square}
\chi^2 = -2\mathcal{L}_0 - 2\sum_n \ln \left[
\alpha \left(\frac{dN_{\mathrm{th}}}{d\varepsilon_p}\right)_n
+ \beta \left(\frac{dN_{\mathrm{bg}}}{d\varepsilon_p}\right)_n ~\right]
\nonumber \\
+ 2 \left(\alpha N_{\mathrm{th}} + \beta N_{\mathrm{bg}} \right)
+ \left(\frac{\alpha-1}{\sigma_{\mathrm{th}}}\right)^2
+ \left(\frac{\beta-1}{\sigma_{\mathrm{bg}}}\right)^2 ,~
\end{eqnarray}
where $-\mathcal{L}_0 = \ln(N_{\mathrm{obs}}!) +
\ln(\sqrt{2\pi}\sigma_{\mathrm{th}}) +
\ln(\sqrt{2\pi}\sigma_{\mathrm{bg}})$, index $n$ runs over all observed
events and $N_{\mathrm{bg}}$ refers to the total number of background events,
which is set to $17.79$ \cite{kamlandR2Data}. Here $\alpha$ and $\beta$ are
the nuisance parameters which are associated with the fractional uncertainties
in theoretical predictions $\sigma_{\mathrm{th}}$ and in background event
counts $\sigma_{\mathrm{bg}}$ respectively.

The KamLAND data release~\cite{kamlandR2Data} provides integrated fission flux
as a function of binned baseline having a bin width of $25~\mathrm{km}$. In our
analysis, we consider baseline length $L$ from the midpoints of each bin having
an error of $\Delta L=12.5~\mathrm{km}$. Consequently, we consider the
uncertainty in the theoretical predictions to be of the form
$\sigma^2_{\mathrm{th}} = \sigma^2_{\mathrm{sys}} + \sigma^2_{\mathrm{L}}$ where
the systematic uncertainty $\sigma_{\mathrm{sys}} = 0.065$
\cite{kamlandSecond2005}. We estimate the uncertainty $\sigma_{\mathrm{L}}$,
arising due to the uncertainty in baseline length $L$, from the equation
(\ref{SurvivalProbability}), as
\begin{equation}
\sigma_{\mathrm{L}} \approx 2
\langle \left(\frac{1-P_{\mathrm{ee}}}{P_{\mathrm{ee}}}\right) \rangle
\langle f(\Delta\phi_{\mathrm{osc}}) \rangle
\left(\frac{\Delta L}{L}\right) ~,
\end{equation}
where $f(\Delta\phi_{\mathrm{osc}}) = \left(\Delta\phi_{\mathrm{osc}}/2\right)
\cot\left(\Delta\phi_{\mathrm{osc}}/2\right)$. By considering the average
survival probability to be $P_{\mathrm{ee}} = 0.658$ for average baseline
$L = 180~\mathrm{km}$ \cite{kamlandSecond2005}, together with $\langle
f(\Delta\phi_{\mathrm{osc}}) \rangle = \ln 2$, we obtain $\sigma_{\mathrm{L}} =
0.050$.

We consider the fractional uncertainty in the background event counts to be
such that $\sigma^2_{\mathrm{bg}} = \sigma^2_{\mathrm{HL}} +
\sigma^2_{\mathrm{acc}} + \sigma^2_{\mathrm{FN}} + \sigma^2_{\mathrm{CO}}$
where $\sigma_{\mathrm{HL}} N_{\mathrm{bg}} = 0.9$ is the uncertainty for
\mbox{$^8$He/$^9$Li} events, $\sigma_{\mathrm{acc}} N_{\mathrm{bg}} = 0.02$ is
the uncertainty for accidental coincidences, and $\sigma_{\mathrm{FN}}
N_{\mathrm{bg}} = 0.89$ is the bound on the fast neutron background
\cite{kamlandSecond2005}. Further, we consider $\sigma_{\mathrm{CO}}
N_{\mathrm{bg}} = 0.32\times 3.42$, where $3.42$ is the number of
${}^{13}\mathrm{C}(\alpha,n){}^{16}\mathrm{O}$ background events up to
$5.5~\mathrm{MeV}$ cut-off. Above this cut-off background contribution is
allowed to float due to inherent uncertainties, while the contributions from
below the cut-off is constrained to be within $32\%$ \cite{kamlandSecond2005}.
Therefore, the total fractional uncertainty for background events becomes
$\sigma_{\mathrm{bg}} = 0.094$.

In order to verify the consistency of our analysis, we minimize $\chi^2$
for the choice $\am = (\Delta m^2_{12}/2$) and $q=1$. The corresponding best fit
values are obtained as $\Delta m^2_{12}/2 = 3.968 \times 10^{-17} ~
\mathrm{MeV}^2$ and $\sin^2\theta_{12} = 0.315$ along with absolute value of
$\chi^2 = 704.47$. In comparison, KamLAND best-fit parameter values are given as
$\Delta m^2_{12}/2 = 3.972 \times 10^{-17} ~\mathrm{MeV}^2$ and
$\sin^2\theta_{12} = 0.314$ \cite{kamlandR2Data}. It demonstrates excellent
agreement between our analysis and the KamLAND analysis. However, the total
$\chi^2$ value at the minimum, which depends on the definition
(\ref{eq:Chi_Square}), differs from the value quoted by KamLAND
\cite{kamlandR2Data}.

\subsubsection{Massless neutrinos with $q \ne 1$}
\label{subsec:Our_no_mass}

\begin{figure}[htbp]
\centering
\includegraphics[height=5.9cm,width=\linewidth]{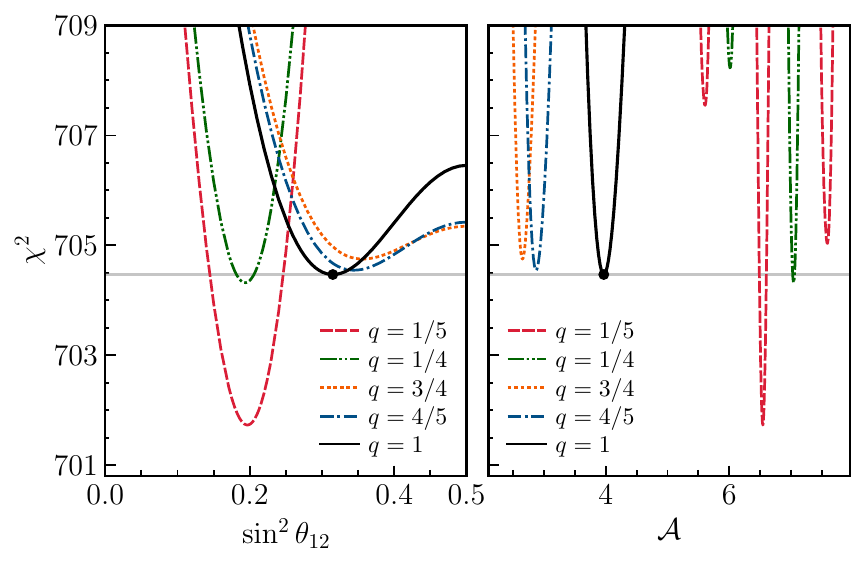}
\caption{The left panel shows the dependence of $\chi^2$ on the parameter
$\sin^2\theta_{12}$ while $\am$ is kept fixed to its best-fit value, given in
Table~\ref{tab:BestFitValues2}. Similarly, the right panel shows its dependence
on $\am$ while $\sin^2\theta_{12}$ is kept fixed to its best-fit value.}
\label{fig:chi2_p2_slices}
\end{figure}

\begin{figure}[htbp]
\centering
\includegraphics[width=\linewidth]{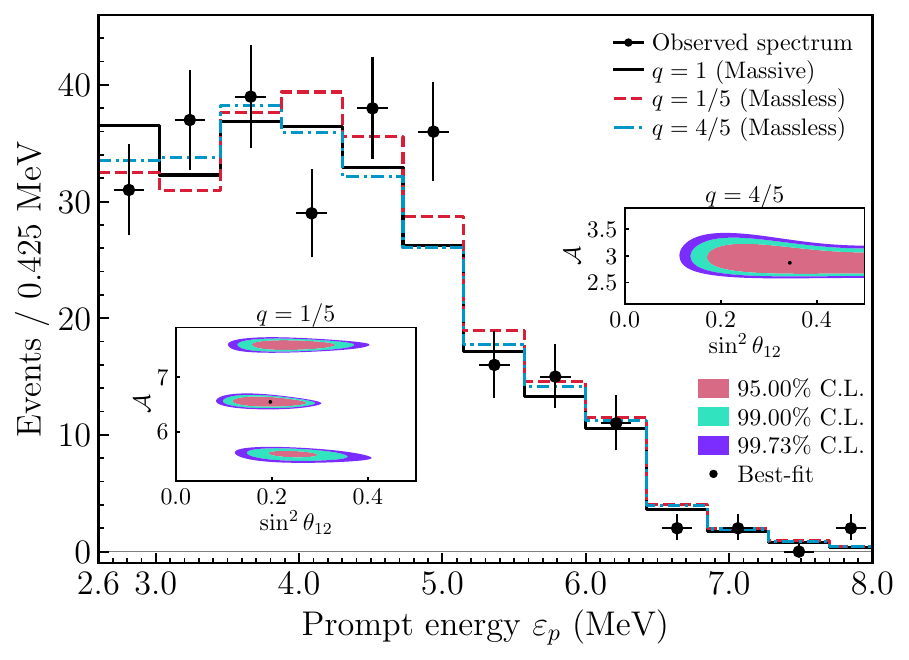}
\caption{The observed event spectrum, with bin width $0.425$ MeV, compared with
the expected number of events, including background events, computed using
best-fit parameters as listed in Table \ref{tab:BestFitValues2}. The confidence
level contours for best-fit parameters are marked as C.L.}
\label{fig:event_spectrum}
\end{figure}

In this section we consider $\am$ and $\sin^2\theta_{12}$ as two free parameters
to be determined through $\chi^2$ minimization for a set of fixed values of the
parameter $q$. The corresponding best-fit parameters for different choices of
$q$ are summarized in Table~\ref{tab:BestFitValues2}. We note that the $\chi^2$
value for $q=1/5$ is \emph{smaller} than the case of massive neutrinos with
$q=1$, whereas for other listed values of $q$, $\chi^2$ values are comparable.
The parameter dependence of $\chi^2$ (\ref{eq:Chi_Square}), around their
best-fit values, is shown in Fig.~\ref{fig:chi2_p2_slices}.
\begin{table}[htbp]
\caption{The best-fit values of the parameters $\am$ and $\sin^2\theta_{12}$
for different but fixed choices of $q$ using \emph{unbinned} analysis.
$\chi^2_{BC}$ refers to Baker-Cousins goodness of fit for the \emph{binned}
spectrum (Fig.~\ref{fig:event_spectrum}) where $\chi^2_{sat} = 53.62$ for
saturated bins.
}
\label{tab:BestFitValues2}
\begin{ruledtabular}
\begin{tabular}{ccccc}
$q$ & $\am ~(10^{-17}\textrm{MeV}^2)$ & $\sin^2\theta_{12}$ & $\chi^2$ &
$\chi^2_{BC}$ \\
\hline
$1/5$  & $6.543$ & 0.197 & 701.73 & 12.59 \\
$1/4$  & $7.041$ & 0.194 & 704.32 & 15.73 \\
$3/4$  & $2.652$ & 0.356 & 704.75 & 12.70 \\
$4/5$  & $2.872$ & 0.344 & 704.55 & 12.67 \\
$1$    & $3.968$ & 0.315 & 704.47 & 13.53 \\
\end{tabular}
\end{ruledtabular}
\end{table}
By integrating the expected event rate, $ \alpha
(dN_{\mathrm{th}}/d\varepsilon_p) + \beta (dN_{\mathrm{bg}}/d\varepsilon_p)$,
over each energy bin, we compute the best-fit \emph{binned} event spectrum
(Fig.~\ref{fig:event_spectrum}). The corresponding Baker-Cousins goodness of fit
values $\chi^2_{BC}$ are listed in Table \ref{tab:BestFitValues2} for different
$q$ values. As implied by the $\chi^2_{BC}$ values, the event spectrum with
$q=1/5$ also shows a better fit with massless neutrinos in comparison to the
event spectrum with massive neutrinos ($q=1$).

\subsection{Discussions}

In summary, we have shown that the flavor states of massless neutrinos, when
viewed as two-level quantum states with energy-dependent level splitting,
result in a higher-likelihood fit to the KamLAND dataset than the
standard massive-neutrino paradigm. Further, by analogy with modified
Jaynes-Cummings models, we have indicated that such level splitting could arise
from the non-trivial spin-gravity coupling in curved spacetime in addition to
the electroweak interaction with matter. In light of the recent results from the
JUNO experiment \cite{JUNO2026first} which features finer energy resolution, it
would be insightful to confront this energy-dependent level splitting framework
with higher-precision data. Finally, further studies are warranted for a deeper
understanding of the level splitting mechanism.

\begin{acknowledgments}
We acknowledge the usage of Kepler cluster of DPS, IISER Kolkata. \emph{Data
Availability:} We thank the KamLAND Collaboration for making their data publicly
available \cite{kamlandR2Data}. Two different sets of source code, written
independently by each author for cross-verification of the results, are publicly
hosted at \cite{GMHPKCode}.
\end{acknowledgments}


%

\end{document}